\documentclass[preprint,prl,preprintnumbers,superscriptaddress]{revtex4}
\usepackage{latexsym}
\usepackage{amsmath}
\usepackage{graphicx}
\usepackage{bm}
\usepackage{epsf}
\usepackage{epsfig}
\usepackage{dcolumn}

\begin{document}

\title{Interlayer binding energy of graphite -- A direct experimental
determination}
\author{Ze Liu}
\affiliation{Department of Engineering Mechanics and Center for Nano and Micro Mechanics,
Tsinghua University, Beijing 100084, China}
\author{Jefferson Zhe Liu}
\affiliation{Department of Mechanical and Aerospace Engineering, Monash University,
Clayton, VIC 3800, Australia}
\author{Yao Cheng}
\affiliation{Department of Engineering Physics, Tsinghua University, Beijing 100084, China}
\author{Quanshui Zheng}
\affiliation{Department of Engineering Mechanics and Center for Nano and Micro Mechanics,
Tsinghua University, Beijing 100084, China}
\affiliation{Institute of Advanced Study, Nanchang University, Nanchang, China}
\email{zhengqs@tsinghua.edu.cn}

\email{zhe.liu@monash.edu}

\begin{abstract}
Despite interlayer binding energy is one of the most important material
properties for graphite, there is still lacking report on its direct
experimental determination. In this paper, we present a novel experimental
method to directly measure the interlayer binding energy of highly oriented
pyrolytic graphite (HOPG). The obtained values of the binding energy are
0.27($\pm $0.02)J/m$^{2}$, which can serve as a benchmark for other
theoretical and experimental works.
\end{abstract}

\pacs{Valid PACS appear here}
\date{\today}
\maketitle


Since the successful fabrication of monolayer graphene from highly oriented
pyrolytic graphite (HOPG) in 2004 \cite{1}, great interests have been
attracted by this perfect two-dimensional crystal made of carbon atoms. The
superior electronic, thermal, and mechanical properties, and the large
specific surface area make it a promising component in next generation of
electronic devices, energy storage and conversion devices, water treatment
application, and the smart (bio) materials \cite{2,3,4,5}. Its inherent
single layer structure determines that the application of graphene, to a
large extent, rely on the detailed understanding and control of it interacts
with its surroundings \cite{6}. However, the nature and strength of
interlayer binding in graphitic materials are poorly understood, despite
that the binding strength in graphite/graphene is relevant to many key
applications, such as graphene electronic devices fabricated on various
substrates, graphite intercalation compounds in Li battery, carbon based
system for hydrogen storage, and graphene based supercapacitors.

Experimentally, there is still lacking report on direct measurements of
graphite interlayer binding energy (BE) and exfoliation energy (EE, the
energy required to remove one graphene layer from a single-crystalline
graphite) \cite{7}. Although there are few reported values obtained
indirectly, those values are quite discrepant. Based on the heat of wetting
data, Girifalco \textit{et al.} \cite{8} obtained the EE value as 43$\pm $5
meV/atom (or 0.26$\pm $0.03J/m$^{2})$. By assuming the simple Lennard-Jones
potential, they further estimated the BE to be about 18{\%} larger than the
EE, while the exact difference remains unknown. Benedict \textit{et al.}
\cite{9} extrapolated the BE as $35_{-10}^{+15} $meV/atom (or $%
0.21_{-0.06}^{+0.09} $J/m$^{2})$ from the measurement on the collapsed
carbon nanotubes. More recently, Zacharia \textit{et al.} \cite{10}
performed desorption experiments of aromatic molecules from a graphite
surface and obtained the approximate graphite EE = 52$\pm $5 meV/atom (or
0.32$\pm $0.03J/m$^{2})$, which yields an estimate of the BE $\sim $ 0.37 J/m%
$^{2}$.

Theoretically, modeling BE as well as EE for graphite is still a question
mark due to the weak interlayer binding involving van der Waals interaction,
which remains a notorious difficult to describe within the standard density
functional theory (DFT) \cite{7,11}. The standard approximations used in
DFT, such as local density approximation (LDA) and generalized gradient
approximation (GGA), cannot accurately describe the long distance
interaction such as van der Waals interactions. LDA calculations lead to
good interlayer distance (e.g., 0.33nm) but binding energy is predicted as
low as 24meV/atom (or 0.15 J/m$^{2}$) \cite{12}. An alternative is the van
der Waals density functional method, which is developed to account the
long-range part of the interactions by using an explicit nonlocal functional
of density. However, the predicted interlayer distance is substantially
higher than experimental value (e.g., 0.36, 0.376, vs. 0.334nm), the elastic
modulus C$_{33}$ is significantly lower (e.g., $\sim $13, 27 vs. 36GPa), and
the binding energy is very scattered (e.g., 24meV/atom (or 0.15 J/m$^{2})$,
45.5 meV/atom (or 0.28 J/m$^{2})$, and 50 meV/atom (or 0.30 J/m$^{2}$)) \cite%
{12,13,14}. Until recently, two comprehensive first-principles calculations
have been carried out. Spanu \textit{et al.} \cite{7} has employed the
quantum Monte Carlo method to obtain the binding energy about 56meV/atom (or
0.34 J/m$^{2})$, and Lebegue \textit{et al.} \cite{11} has used the
adiabatic-connection fluctuation-dissipation theorem in the direct random
approximation to obtain the binding energy as 48meV/atom (or 0.29 J/m$^{2})$.

The above brief review shows a clear need of directly experimental
determination of the BE for graphite. In this paper, we introduce a novel
method to directly measure BE, which is motivated by our recent discovery of
self-retraction motion phenomenon of micrometer graphite/SiO$_{2}$ flakes on
graphite islands \cite{15}. The idea is to assemble a thin graphite flake to
span over a graphite step. As illustrated in Fig.1(a), the considered
graphite flake consists of multilayered graphenes assembled in AB-stack, and
the graphite step is single-crystalline with atomically smooth graphene top
surfaces. The key of this experimental technique is to make the contact
areas (C-D and E-F indicated in Fig. 1(a)) between the flake and the step in
the AB-stack assembling. In our experiments, each graphite flake was coated
on its top surface with a SiO$_{2}$ thin film. The deformation energy $U$ of
the above of graphite/SiO$_{2}$ flake and graphite step can be modeled as a
function of the step height $\Delta $, the span length $L$, the thicknesses $%
h_{s}$ and $h_{g}$ of the SiO$_{2}$ film and the graphite flake, and the
material elastic constants. Since a virtual increment $\delta L$ of length $%
L $ leads to the increase of the graphite surface area 2$b\times \delta L$
and, consequently, the increment of the total surface energy 2$\gamma
b\times \delta L$, where $b$ denotes the flake width and $\gamma $ is the
graphite basal plane surface energy, we obtain the equilibrium equation of
the above system in the following form:

\begin{equation}  \label{eq1}
2\gamma =-\frac{1}{b}\frac{\partial U}{\partial L}.
\end{equation}
Using the above-proposed method, we measured the interlayer binding energy,
namely 2$\gamma $, of HOPG to be 0.27$\pm $0.02 J/m$^{2}$ (or 44$\pm $3
meV/atom). The details of our experiments are reported below.

The HOPG samples were purchased from Veeco (ZYH grade). As illustrated in
Fig. 2(a), graphite islands were fabricated by using the same technique as
that reported in Ref. \cite{16}. After mechanical exfoliation, a clean and
fresh top surface of the HOPG sample was obtained \cite{17}. A silicon
dioxide (SiO$_{2})$ film was then grown on the sample top surface by plasma
enhanced chemical vapor deposition (PECVD), and then followed by electron
beam lithography and reaction ion etching. Figure 2(b) shows the scanning
electron microscopy (SEM, FEI Quanta 200F) image of an obtained island with
typical length 2-5$\mu $m and height 400nm. Figure 2(c) is the SEM image of
the side-view of an island with the top flake tilted 30 degrees.

Similar to Ref. \cite{15}, we employed a micromanipulator MM3A (Kleindiek)
that had been set in a SEM (FEI Quanta 200F) to perform our experiments
under an ultrahigh vacuum condition (1.19 - 6.72$\times $10$^{-4}$ Pa) and
the room temperature. Electron beam with 30 KV acceleration and 3 nm
spot-size was used to monitor the fabrication process. The \textit{in situ}
process can be seen from Supplementary Movie. Schematically we show the
experimental process in Fig. 3. We placed the MM3A tip on the top surface of
a selected island and then pushed it in a lateral direction (Fig. 3(a)). A
graphite/SiO$_{2}$ flake was then sheared out from its platform (base flake
of the graphite island). The flake was found to be fully self-retractable
after removing the tip. Our study on the self-retraction mechanism has
revealed that the slipping plane corresponds to a boundary between two
single-crystalline graphite grains and the original assembling between the
moved flake and platform (before shear) is not AB-stacked, which leads to
the large-scale superlubricity \cite{15, 18} and consequently the
self-retraction motion. To prevent the self-retraction, we used the tip to
rotate the flake for a certain angle until the flake was suddenly locked up,
which corresponds to an AB-stack assembling as revealed in our recent study
\cite{18}. In the experiment, pushing the locked top flake again in the
opposite lateral direction leads to the separation of the flake into two
parts: the top and the middle flakes. The observed self-retraction between
these two separated flakes (See Supplementary Movie) indicates once more a
non AB-stack contact. To prevent the self-retraction, we rotated the top
flake with respect to middle flake to another `lock-up' state (Fig. 3(d)),
and finally two flakes over the platform are both locked-up.

The optical microscope (OM, HiRox KH-3000) image of a typical locked-up
example in our experiments is shown in Fig. 3(e), where the top flake (blue
colored) spanned over the middle graphite flake. Fig 3(f) shows the height
profile along the black line in Fig. 3(e), obtained by using an atomic force
microscope (AFM). The measured step height or the middle flake thickness is $%
\Delta $* = 53.7$\pm $0.9 nm. In comparison, the height drop along the top
flake surface was measured as $\Delta $ = 53.8$\pm $2.7 nm, which is almost
the same as the thickness of the middle flake. Such a good agreement
confirms the good adhesion of the top flake with the middle flake and
platform. This negligible difference could come from the surface roughness
of the SiO$_{2}$ thin film. The good adhesion is also supported by colour
image in Fig. 3(e), where the blue color is fully reproduced at both flat
sides of the top flake.

Figure 4(a) shows the AFM measured height profiles along the lines in the
insert for the sample prepared in SEM as explained before. The thickness of
the top flake is measured as $h$ = 325$\pm $1nm and the height of the step
(i.e., the thickness of the middle flake) is $\Delta $ = 30$\pm $3 nm. By
using MM3A micromanipulator to take the top flake off its platform and then
stand the flake up, we measured the thickness of the SiO$_{2}$ film as 205$%
\pm $8 nm (Fig. 4(b)). Using the measured values $\Delta $ = 30nm, $h_{s}$ =
205nm, $h_{g}$ = 120nm, and well-accepted elastic constants of SiO$_{2}$ and
graphite in our finite element (FEM) model (see details in Supplementary
Information), we obtained the deflection curve $y(x$, $L)$ of the top
surface of the model graphite/SiO$_{2}$ flake at different span length $L$.
Finally, we can quite accurately determine the span length $L$ of each AFM
measured height profile by least square fitting it to the $y(x$,$L)$ results
obtained from our FEM analysis. The actual $L$ in our experiments should
correspond to the $L$ value with the least fitting error. One example of
such fitting is given in Fig. 4(c) with the fitted span lengths $L$ = 890
nm. The error bound of $L$ is estimated smaller than 20nm.

It should be noted that expressing the elastic strain energy $U$ in Eq. (\ref%
{eq1}) by an analytic model is rather difficult due to the following two
aspects. First, the bulk graphite has the highest elastic anisotropy \cite%
{19} : the highest in-plane elastic modulus (e.g., $\sim $1000GPa) and very
weak interlayer shear modulus (e.g., $\sim $4.5GPa). The shear deformation
near the interface between SiO$_{2}$ film and graphite flake is expected to
be significant and difficult to theoretically model. Second, the elastic
deformation of the step (made up by the middle flake and the platform) is
difficult to describe as well, although it turns out to be important. Finite
element method (FEM) analyses were thus employed to calculate the strain
energy $U$. The details of FEM model are included in Supplementary
Information (SI). Figure 1(b) plots the calculated strain energies $U(L)$
and Fig. 1(c) gives the approximations of BE through following central
finite difference of Eq. (\ref{eq1}):
\begin{equation}
2\gamma (L)\approx -\frac{U(L+\delta L)-U(L-\delta L)}{2b\times \delta L},
\label{eq2}
\end{equation}%
with $\delta L$ = 20 nm. Finally, the binding energy is approximated as 0.27
$\pm $ 0.02 J/m$^{2}$, corresponding to determined length $L$ = 890 $\pm $
20 nm (Fig. 1 (c)).

To test the reliability of the above-proposed method, we measured the second
island sample with different size and shape. The measured binding energy is
0.30 $\pm $ 0.025 J/m$^{2}$, which agrees pretty well with the first sample
shown in Figure 4(a). Details are given in SI.

It is well known that chemisorption, physisorption, and insertion of gases
inside graphite can drastically change the surface/interface properties \cite%
{20}. The binding energy of graphite under vacuum was estimated about 100
times higher those in an environment with air, oxygen or water vapor \cite%
{21}. To avoid such artificial effects, we note again that above-reported
experiments are carried out in SEM with ultra-high vacuum condition. Similar
experiments but in atmospheric conditions were also done using an MM3A
assisted by an optical microscope. The binding energy measured is just a
little smaller than that from SEM, $\sim $ 0.22J/m$^{2}$. We believe that
the self-retraction motion of graphite/SiO$_{2}$ flakes can self-clean the
adsorbate on the exposed sliding surfaces \cite{22}.

In summary, a novel experimental method is presented here to directly
measure the interlayer binding energy of highly oriented pyrolytic graphite:
2$\gamma $ = 0.27 $\pm $ 0.02 J/m$^{2}$ (or 44 $\pm $3 meV/atom). The error
bar is mainly caused by the roughness of the top SiO$_{2}$ thin film. It can
be improved significantly by either extracting a smoother deflection profile
from comparison of the AFM height profiles along lines on top flakes before
and after its spanning over the step, or simply using other types of top
thin films and improved ion etching process. Our proposed method can be
easily extended to measure the binding energies between the
graphite/graphene and other types of substrates. It can also be used in
other systems, particularly lamellar materials and thin films. Considering
the difficulty in measuring the interface binding energy in
micro/nano-materials, our method could serve as a general solution.

Q.S.Z. acknowledges the financial support from NSFC through Grant No.
10832005, the National Basic Research Program of China (Grant No.
2007CB936803), and the National 863 Project (Grant No. 2008AA03Z302). J.Z.L.
acknowledges new staff grant 2010 and small grant 2011 from engineering
faculty of Monash University.

\pagebreak

\begin{figure}[htbp]
\centerline{\includegraphics[width=0.8\textwidth]{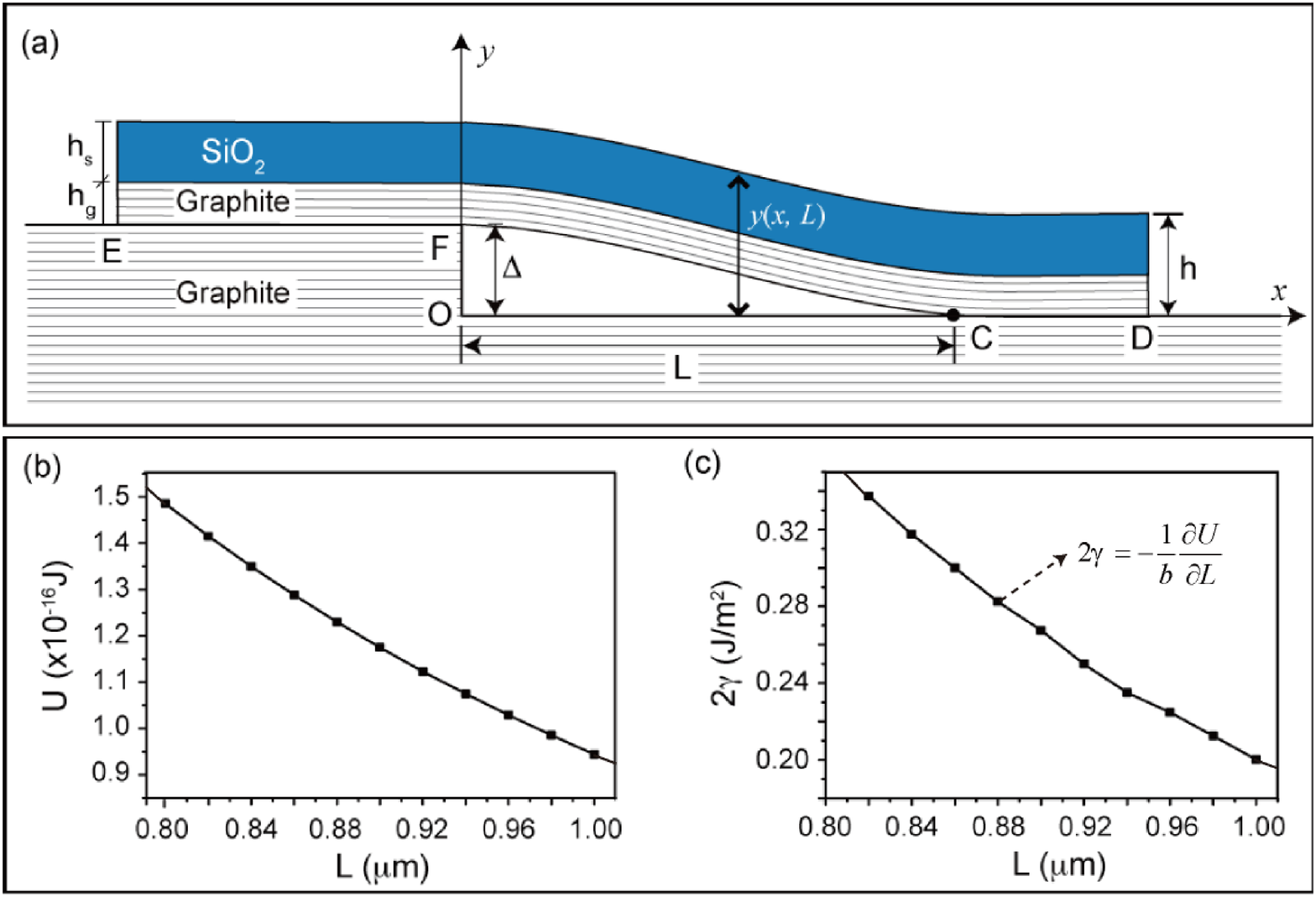}}
\caption{(color online) (a) Sketch image of a graphite/SiO$_{2}$ flake
spanned a graphite step of height $\Delta $ with spanned length $L$. (b) The
system strain energy (black square dots) versus spanned length obtained by
using ABAQUS simulations, where the width is set as a unit for plane strain
and the other parameters ($\Delta $ = 30nm, $h_{s }$= 205nm, $h_{g }$=
120nm) are the measured values using AFM and SEM, see text for details. (c).
The binding energy obtained by central finite difference of the data in (b) }
\label{fig1}
\end{figure}

\pagebreak

\begin{figure}[htbp]
\centerline{\includegraphics[width=0.8\textwidth]{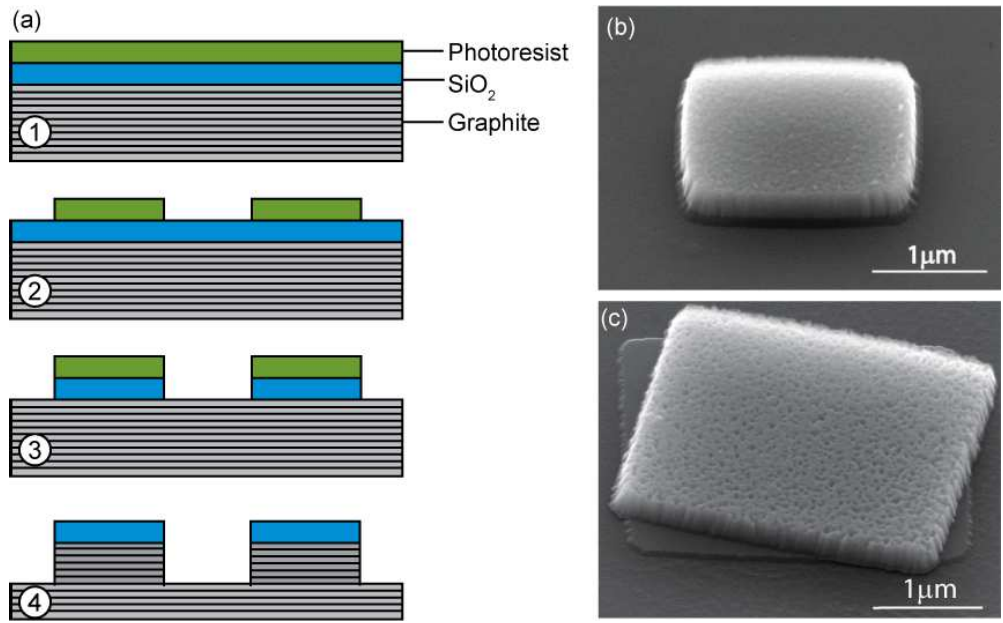}}
\caption{(color online) (a) Fabrication illustration of the graphite
islands. SiO$_{2}$ thin film is grown on top of the island by using plasma
enhanced chemical vapor deposition. Electron beam lithography and reaction
ion etching is then used to fabricate the island. (b)(c) Scanning electron
microscopy image of the side-view of the fabricated island as the sample
stage tilted 30 degrees.}
\label{fig2}
\end{figure}

\pagebreak

\begin{figure}[htbp]
\centerline{\includegraphics[width=0.8\textwidth]{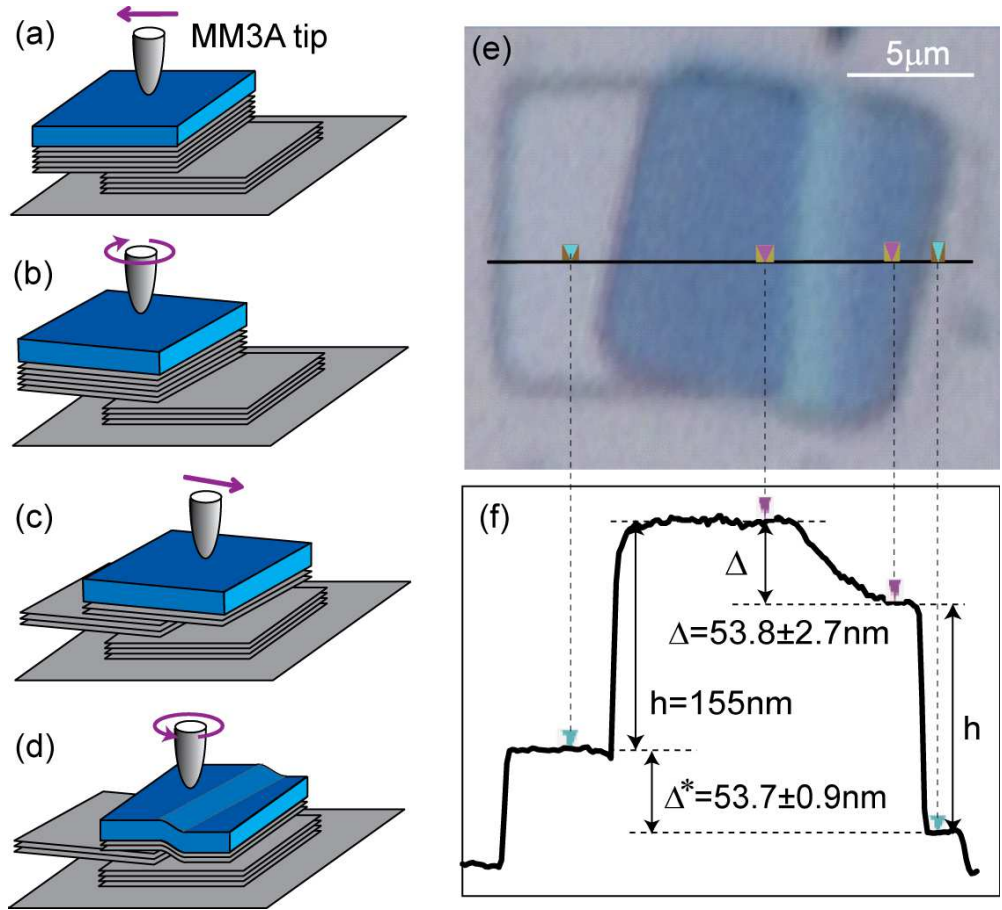}}
\caption{(color online) Illustration of the experimental process.
(a) An MM3A tip is used to slide a graphite/SiO$_{2}$ flake with
respect to the bottom flake. (b) The graphite/SiO$_{2}$ flake is
then rotated by a certain angle to a `lock-up' state (See text and
Supplementary Movie). (c) After that the tip is moving in the
opposite direction to split the slided flake into two parts: top and
middle flake. (d) The top flake is spanned over the middle and
bottom flakes and then it is rotated by the MM3A tip to another
`lock-up' state. (e) The optical microscope image of one obtained
sample. (f) The profile obtained by atomic force microscopy (AFM)
indicates that the adhesion is indeed taken place. The blue color in
(e) reveals the constructive interference of $\sim \protect\lambda
$/2 (130nm) thickness of SiO$_{2}$ layer, that is consistent with
the measurement in (f) The appearance of a greenish blue band in (e)
reveals that the slope surface is quite uniform in width direction.}
\label{fig3}
\end{figure}

\pagebreak

\begin{figure}[htbp]
\centerline{\includegraphics[width=0.8\textwidth]{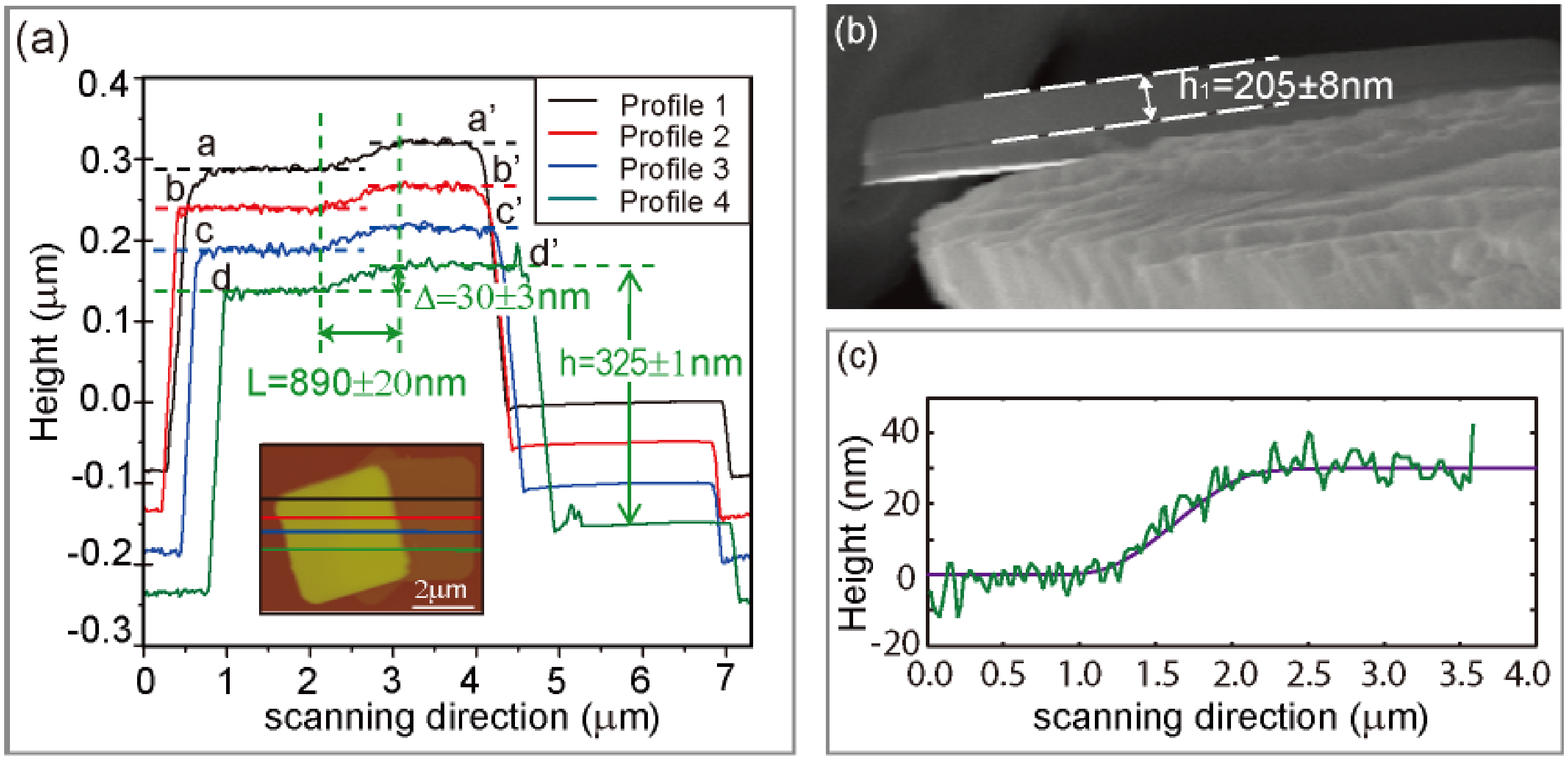}}
\caption{(color online) AFM height profiles along the colored lines in Inset
of a double locked-up sample prepared in a SEM in a high vacuum condition.
These profiles are shifted for visulization clarity. (b) A graphite/SiO$_{2}$
flake adhered to a tip (bottom), showing the tickness of the SiO$_{2}$ film
to be 205$\pm $8 nm. (c) One representative least square fittings of
deflection curves, yielding a consistent span length $L$ = 890$\pm $20nm for
each of the four profiles.}
\label{fig4}
\end{figure}


\begin{thebibliography}{99}
\bibitem{1} K. S. Novoselov et al., Science 306, 666 (2004).

\bibitem{2} A. K. Geim, Science 324, 1530 (2009).

\bibitem{3} A. K. Geim, et.al., Nat. Mater. 6, 183 (2007).

\bibitem{4} M. D. Stoller, et.al., Nano Lett., 8, 10 (2008).

\bibitem{5} N. Mohanty, et.al., Nano Lett., 8, 12 (2008).

\bibitem{6} X. Du, et.al., Nat. Nanotechnology 3, 491(2008).

\bibitem{7} L. Spanu, et.al., Phys. Rev. Lett. 103, 196401 (2009).

\bibitem{8} L.A. Girifalco, et.al., J. Chem. Phys. 25, 693(1956).

\bibitem{9} L. Benedict et al., Chem. Phys. Lett. 286, 490 (1998).

\bibitem{10} R. Zacharia, et.al., Phys. Rev. B 69, 155406 (2004).

\bibitem{11} S. Lebegue, et.al., Phys. Rev. Lett. 105, 196401 (2010).

\bibitem{12} M. Dion, et.al., Phys. Rev. Lett. 92, 126402 (2004).

\bibitem{13} E. Ziambaras, et.al., Phys. Rev. B 76, 155425 (2007).

\bibitem{14} S. D. Chakarova-Kack, et.al., Phys. Rev. Lett. 96, 146107
(2006).

\bibitem{15} Q.S. Zheng, et.al., Phys. Rev. Lett. 100, 067205 (2008).

\bibitem{16} X.K. Lu et al., Nanotechnology 10, 269 (1999).

\bibitem{17} Z. Liu, et.al., Appl. Phys. Lett. 96, 201909 (2010).

\bibitem{18} Z. Liu et al., Self-retracting motion of graphite micro-flakes:
superlubricity in micrometer scale (submitted, see also:)

\bibitem{19} L. F. Wang, et.al., Appl. Phys. Lett. 90, 153113 (2007).

\bibitem{20} H. Zaidi, et.al., Thin Solid Films, 264, 46 (1995).

\bibitem{21} P. J. Bryant, et.al., Mechanics of Solid Friction (Elsevier,
Amsterdam, 1964).

\bibitem{22} Z. Liu, et.al., Graphite nanoeraser, Nanotechnology (revised
version under review, see also: http://arxiv.org/abs/1010.4102).

\end{thebibliography}
\end{document}